\documentclass[useAMS,usenatbib]{mn2e}
\usepackage{graphicx}
\usepackage{color}
\usepackage{aas_macros}

\outer\def\gtae {$\buildrel {\lower3pt\hbox{$>$}} \over 
{\lower2pt\hbox{$\sim$}} $}
\outer\def\ltae {$\buildrel {\lower3pt\hbox{$<$}} \over 
{\lower2pt\hbox{$\sim$}} $}

\newcommand{\kep} {\sl Kepler}
\newcommand{\src} {KIC 10449976}

\begin{document}

\title[An extreme-helium subdwarf in the {\sl Kepler} field]
{KIC 10449976: discovery of an extreme-helium subdwarf in the {\sl Kepler} field}

\author[C.~S.~Jeffery, G.~Ramsay, Naslim, N., et al.]{C.~S.~Jeffery$^{1}$\thanks{E-mail: csj@arm.ac.uk}, 
G.~Ramsay$^{1}$,  N. Naslim$^{1}$,  R. Carrera$^{2,3}$, S. Greiss$^{4}$,
T. Barclay$^{5,6}$, \and R. Karjalainen$^{7}$, A. Brooks$^{1,8}$, P. Hakala$^{9}$ \\
$^{1}$Armagh Observatory, College Hill, Armagh, BT61 9DG\\
$^{2}$Instituto de Astrof\'isica de Canarias, La Laguna E-3200, Tenerife, Spain\\
$^{3}$Departamento de Astrof\'isica, Universidad de La Laguna, La Laguna E-38205, Tenerife, Spain\\
$^{4}$Department of Physics, University of Warwick, Coventry, CV4 7AL\\
$^{5}$NASA Ames Research Center, M/S 244-40, Moffett Field, CA 94035, USA\\ 
$^{6}$Bay Area Environmental Research Institute, Inc., 560 Third St. West, Sonoma, CA 95476, USA\\
$^{7}$Isaac Newton Group of Telescopes, Apartado de Correos 321, E-38700 Santa Cruz de la Palma, 
Canary Islands, Spain\\
$^{8}$Mullard Space Science Laboratory, University College London, Holmbury St. Mary, Dorking, Surrey RH5 6NT\\
$^{9}$Finnish Centre for Astronomy with ESO (FINCA) , University of Turku,
V\"{a}is\"{a}l \"{a}ntie 20, FI-21500 PIIKKI\"{O}, Finland\\
}

\date{Accepted 2012 December 6.  Received 2012 December 6; in original form 2012
November 8}

\pagerange{\pageref{firstpage}--\pageref{lastpage}} \pubyear{2011}

\maketitle

\label{firstpage}

\begin{abstract}
Optical spectroscopy of the blue star {\src} shows that it is an extremely 
helium-rich subdwarf with effective
temperature $T_{\rm eff}=40\,000\pm300$\,K and surface gravity $\log
g=5.3\pm0.1$.  Radial-velocity measurements over a
five-day timescale show an upper variability limit of $\approx50\pm20$ km
s$^{-1}$. {\kep}  photometry of {\src} in both long and short
cadence modes shows evidence for a periodic modulation on a
timescale of $\approx3.9$ days. We have examined the possibility that
this modulation is not astrophysical but conclude it is most likely
real. We discuss whether the modulation could be caused by a low-mass
companion, by stellar pulsations, or by spots. The identification of
any one of these as cause has important consequences for understanding
the origin of helium-rich subdwarfs.
\end{abstract}


\section{Introduction}

Prior to the launch of NASA's {\kep} satellite, an extensive programme
to identify bright dwarf G and K stars with minimal stellar activity
was carried out, which resulted in the {\sl Kepler Input Catalog}
(KIC) \citep{brown11}.  Although a small number of photometric
variability surveys were carried out pre-launch
\citep[]{hartman04,pigulski09,feldmeier11} they were either not
especially deep, did not have a wide sky coverage or did not have
a cadence shorter than a few minutes. To identify short-period
variable sources which would be interesting to be observed using
{\kep} in Short Cadence (SC) mode, the {\sl RATS-Kepler} project
commenced in the summer of 2011 (Brooks et al., in prep). As part of a
follow-up programme to determine the nature of sources which showed
short period variability and/or were blue, we obtained
medium-resolution spectroscopy of a sample of objects using the Isaac
Newton Telescope (INT) in La Palma. One of the targets, KIC 10449976
($\alpha$=18h 47m 14.1s, $\delta$=+47$^{\circ}$ 41$^{'}$ 46.9$^{''}$:
J2000, $g$=14.49), was observed because it was clearly blue
($g-r=-0.5$, KIC). The optical spectrum was notable because it showed
a number of strong lines due to neutral helium and weak or absent
hydrogen Balmer lines.

Early-type stars with absorption spectra dominated by neutral  or ionized 
helium are extremely rare and fall into one of
(roughly) two classes. Slightly better known are the extreme helium
stars: low-gravity stars of spectral types B and A
\citep{jeffery08.hdef3.a}. A few low-gravity helium stars of
spectral-type O are also known. These are almost certainly related to
the cooler R\,Coronae Borealis variables, and are currently considered
to be the product of the merger of a helium white dwarf with a
carbon-oxygen white dwarf \citep{jeffery11a}.  Extremely helium-rich
hot subdwarfs are found (naturally) at higher gravities with early-B
or late-O type spectra \citep{naslim10}.  Again, the merger of two
helium white dwarfs is the strongest contender for their production
\citep{zhang12a}.

Many of the extreme helium stars are small-amplitude flux variables,
most probably caused by opacity-driven radial or non-radial pulsations
\citep{saio88b}. However, probably due to extreme non-adiabacity, the
pulsations are relatively irregular and have proved difficult to study
systematically with ground-based photometry \citep{kilkenny99c}.  No
extreme helium star has been discovered since 1986 \citep{drilling86}.

The helium-rich subdwarfs may be subdivided into an extremely
helium-rich class ($n_{\rm He}>80\%$) and an intermediate helium-rich
class ($10\%<n_{\rm He}<80\%$) \citep{naslim12}.  None of the
extremely helium-rich group are known to pulsate or to be members of a
binary system, with two exceptions. V652\,Her \citep{jeffery01b} and
BX\,Cir \citep{woolf01} are somewhat cooler than the remainder and lie
in the Z-bump instability strip, and pulsate with a period of 0.1\,d
\citep{landolt75,kilkenny95}.  Amongst the intermediate group, there
is at least one binary: CPD$-20^{\circ}1123$ \citep{naslim12} and one
pulsator: LS\,IV$-14^{\circ}116$ \citep{ahmad05}. The latter shows an
extraordinary surface chemistry, with 4 dex overabundances of
zirconium, strontium, and yttrium and a 3 dex overabundance of
germanium. This chemistry is probably produced by radiation-dominated
diffusion \citep{naslim11}.

\citet{ostensen10} describes a systematic survey  for compact pulsators 
and identifies three  He-sdOB stars in the {\kep} field; 
(Galex) J19034+3841, 
(SDSS) J19352+4555
and J19380+4649.  
Surface properties have not been published.
None pulsates, but J19352+4555
shows irregular low-fequency light variations \citep{ostensen10}.  
The maximum light variations in the other two He-sdOBs at low-frequencies 
(100 -- 500 $\mu$Hz) are 
117 parts per milllion (J19034+3481) and 29 parts per million  (J19380+4649). 
An intermediate helium-rich blue-horizontal-branch star in
the {\kep} field (KIC\,1718290) has been identified as a non-radial
g-mode pulsator \citep{ostensen12}. 

No previously known
extreme-helium stars or helium-rich subdwarf lies in the {\kep}
field, so the discovery of new members of either group lying within
the field offers an opportunity to explore the questions of duplicity
(a close companion makes the white-dwarf merger model difficult to
maintain) and pulsation stability.

In this paper we present an analysis of the spectrum of KIC 10449976
to measure its effective temperature, surface gravity and surface
composition.  We analyse six quarters of {\kep} photometry, which
appear to show some evidence of small-amplitude flux variability. We
present a series of radial-velocity measurements obtained to establish
whether it could be a binary. We discuss possible explanations for the
variability.

\begin{figure}
\begin{center}
\setlength{\unitlength}{1cm}
\begin{picture}(7,5)
\put(-0.5,-0.5){\includegraphics{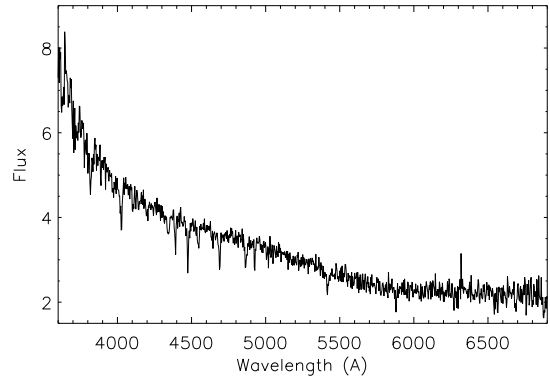}}
\end{picture}
\end{center}
\caption{The flux calibrated optical spectrum of KIC 10449976 taken
  using the IDS on the INT on 2012 June 28.}
\label{f_spec}
\end{figure}

\begin{figure}
\begin{center}
\setlength{\unitlength}{1cm}
\begin{picture}(6,5)
\put(-1,+5){\includegraphics{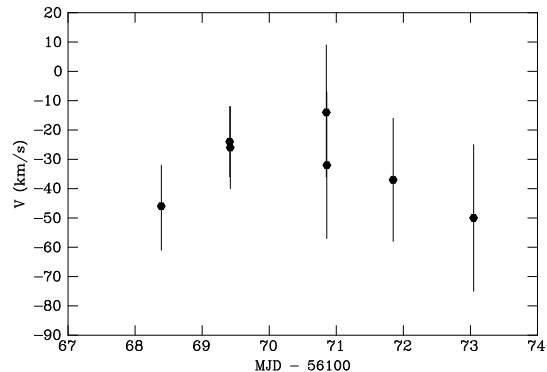}}
\end{picture}
\end{center}
\caption{The heliocentric radial velocity of {\src} as measured in 2012 August and September. 
The error bars represent
  the total error  from the cross-correlation function, the wavelength calibration 
 and the template velocity measurement. }
\label{radial}
\end{figure}

\begin{table*}
\begin{center}
\caption{Summary of spectroscopic observations of {\src}.}
\begin{tabular}{lccccccccc}
\hline
Telescope & Date&\# spectra&Exp. time&Grism&Slit width&FWHM & Velocity  & $\delta V_1^1$ & $\delta V_2^2$ \\
 &   &               &(s)      &     &('')    &(\AA) & (km\,s$^{-1}$) & & \\
\hline
 INT & 20120628 & 2 & 180 & R400V & 1.0 & 4.0 & $-2.9$ & 8.1 & 13.3 \\
 WHT & 20120810 & 3$^3$ & 900 & R600R & 1.2 & 2.0 & $-16.1$ & 10.5 & 12.1 \\
 INT & 20120828 & 1 & 600 & R300V & 0.5 & 5.4 & $-46.3$ & 9.9 & 14.5 \\ 
 INT & 20120829 & 2 & 600 & R300V & 0.5 & 3.9 & $-24.0$ & 6.4 & 12.3 \\
       &                   &   &         &            &       &      &  $-25.8$ & 9.6 & 14.2 \\
 INT & 20120831 & 2 & 600 & R632V & 0.5 & 2.6 & $-13.6$ & 21.5 & 22.3 \\
       &                   &   &         &            &       &      &  $-32.0$ & 9.6 & 14.2 \\
 INT & 20120901 & 1 & 600 & R632V & 0.5 & 1.7 & $-37.0$ & 18.9 & 20.9 \\
 INT & 20120902 & 1 & 600 & R632V & 0.5 & 1.7 & $-49.9$ & 21.8 & 25.1 \\
\hline
\end{tabular}\\
Notes: 1: Formal error in the cross-correlation measurement,\\ 2: Total error including wavelength calibration and template velocity,\\ 3: All three  spectra were combined for the velocity measurement. 
\end{center}
\label{t_spec}
\end{table*}

\section{Optical Spectroscopy}

KIC 10449976 was initially observed using the Intermediate Dispersion
Spectrograph (IDS) on the Isaac Newton Telescope (INT) on the night of
2012 June 28 (see Table \ref{t_spec} for details). The spectra were
reduced using standard procedures with the wavelength calibration
being made using a CuNe+CuAr arc taken shortly after the object
spectrum was taken. A flux standard was observed so that the spectra
could be flux calibrated. The individual spectra were combined to give
a mean spectrum which shows all the characteristics of a
helium-rich star (Figure \ref{f_spec}), including all expected He{\sc
  i} lines, He{\sc ii}\,4686\AA\ and the He{\sc ii} Pickering series.
With He{\sc ii}4541\AA\ slightly weaker than the H$\gamma$+He{\sc
  ii}\,4340\AA\ blend, the hydrogen abundance is evidently small, but
not zero. By comparison with the classification spectra of
\citet{drilling12}, we assign an approximate spectral class of
sdB0VI:He35 (the discovery spectrum is at a lower resolution than the
classification system: 2\AA).  Adopting the Drilling et
al. calibration gives an effective temperature $T_{\rm
  eff}\approx38\,000\pm2\,000\,{\rm K}$, surface gravity $\log
g\approx5.3\pm0.3$ and helium-to-hydrogen ratio $n_{\rm He}/n_{\rm H}
> 10$. This places KIC 10449976 firmly amongst the extremely 
helium-rich subdwarfs, rather than the extreme helium stars. 

We obtained a further series of observations in 2012 August and
September (Table \ref{t_spec}) to search for radial-velocity
variations. We took arc lamps for calibration before and after the
target; this provides an estimate of the uncertainties in the
wavelength calibration.  Radial velocities were obtained by the method
of cross-correlation.  The four INT spectra from the nights of 2012
August 31, September 1 and 2 were coadded to form a template. All of the
spectra, including the template, were rectified, continuum-subtracted,
and converted to log wavelength.  The cross-correlation function (ccf)
between each individual spectrum and the template was computed and
converted to velocity units. The position of the peak of the ccf was
measured by fitting a Gaussian. The radial velocity of the template was
obtained by cross-correlation with the best-fit theoretical spectrum at
rest wavelength (see below). Heliocentric corrections were applied. The results 
 are shown in Fig. \ref{radial} and Table \ref{t_spec}. The measurements 
are consistent with there being no variation in the radial velocity; 
they are also consistent with a variation of up to $50\pm20$km\,s$^{-1}$
over an interval of days. 
 
Spectra were also obtained using the William Herschel Telescope (WHT)
 on 2012 August 10 (see Table \ref{t_spec} for
details) using the Intermediate dispersion Spectrograph and Imaging
System (ISIS). Spectra were reduced in a similar manner to those
obtained using the INT, and the radial velocity measured in the same way.


\begin{figure*}
\begin{center}
\setlength{\unitlength}{1cm}
\begin{picture}(8,8)
\put(-4.5,-0.5){\includegraphics{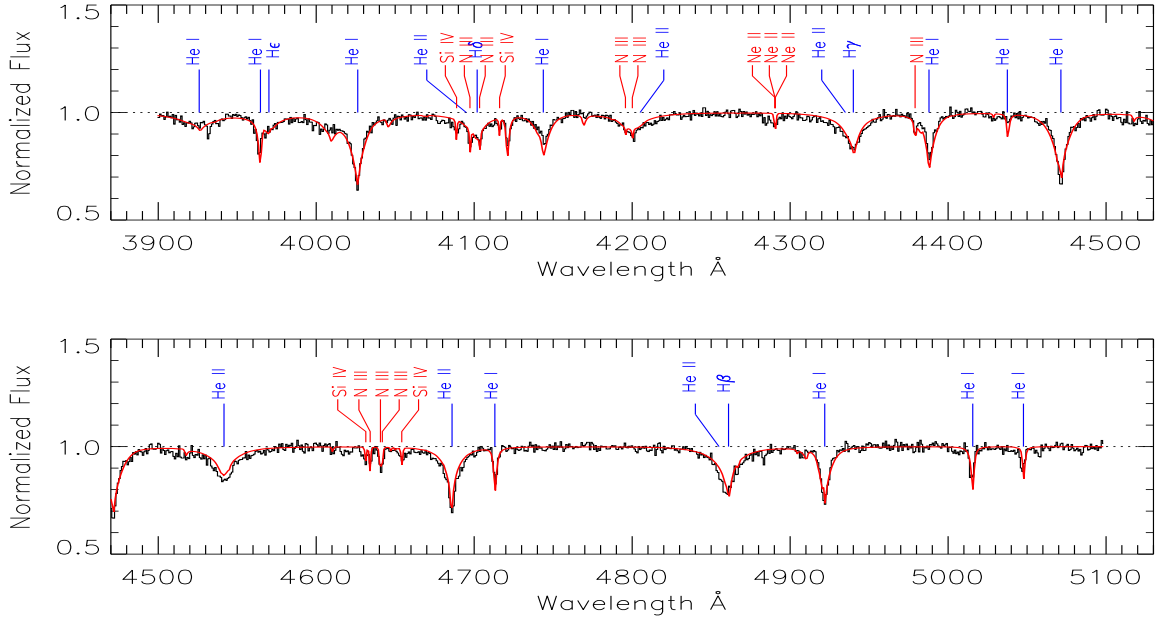}}
\end{picture}
\end{center}
\caption{The normalised WHT optical spectrum of {\src} fitted with the
 model described in \S 3 optimized for abundances of  C, N, Si and Ne, 
 together with line identifications. }
\label{f_wht}
\end{figure*}

\begin{table}
\begin{center}
\caption{Atmospheric parameters of {\src}. Solar abundances are from \citet{grevesse98}; note 
that fractional abundances by number are not conserved when hydrogen is converted to helium. 
}
\begin{tabular}{lllll}

\hline
$T_{\rm eff}$ & 40.0 & $\pm 0.3$ kK \\
$\log g$ & 5.33 & $\pm 0.10$ (cgs) \\
$v_{\rm turb}$ & 5 & km\,s$^{-1}$ & assumed \\[2mm]
$n_{\rm H}$ & 0.06 & $\pm 0.02$ \\
$n_{\rm He}$ & 0.94 & $\pm 0.02$ \\
                    &          &               &  Sun \\
$n_{\rm C}$ & $< 4.6\times10^{-6}$  &    & $9.5\times10^{-4}$ \\
$n_{\rm N}$ & $2.9\times10^{-4}$ & $\pm1.2\times10^{-4}$ & $8.3\times10^{-5}$ \\
$n_{\rm Si}$ & $9.6\times10^{-6}$  & $\pm6.7\times10^{-6}$  & $1.0\times10^{-4}$ \\
$n_{\rm Ne}$ & $9.6\times10^{-4}$  & $\pm5.7\times10^{-4}$  & $1.2\times10^{-4}$ \\
\hline
\end{tabular}
\end{center}
\label{t_params}
\end{table}

\section{Atmospheric Parameters}

Best-fit atmospheric parameters for KIC 10449976 were established by
interpolation in a grid of synthetic spectra computed from a grid of
line-blanketed model atmospheres computed in local thermodynamic,
hydrostatic and radiative equilibrium.  The grid covers a wide range
in effective temperature $T_{\rm eff}$, surface gravity $g$, and
helium abundance $n_{\rm He}$ for a number of distributions of
elements heavier than helium, including solar, 1/10 solar and other
custom-designed mixtures \citep{behara06}.  For the optical spectrum
of KIC 10449976 we sought solutions in the range $32\,000 < T_{\rm
  eff} < 50\,000$\,K, $4.5 < \log g < 6.0$, and $0.90 < n_{\rm He} <
0.999$ over the wavelength interval $3750 - 5100$\AA\ ($n$ is the
fractional abundance by number).  Solutions were obtained by
$\chi^2$-minimisation to the continuum-rectified and normalised
spectrum using the optimisation code SFIT \citep{jeffery01b}. With no
indicators for microturbulent velocity at the spectral resolutions
used we assumed a value 5\,km\,s$^{-1}$. The procedure was applied
first to the June INT spectrum; this established KIC 10449976 to be,
indeed, a helium-rich subdwarf. It was subsequently applied to the
higher-resolution higher-signal WHT spectrum.  A 1/10 solar mixture
was chosen in preference to a solar mixture on the basis of the
strengths of visible metal lines. Virtually identical results were
obtained in both cases.  
The atmospheric parameters are shown
in Table~\ref{t_params}. 

The fit to the WHT spectrum (Figure \ref{f_wht}) showed that a number
of significant strong absorption lines could be resolved, in addition
to those due to hydrogen and helium.  Notably, these included N{\sc
  iii} lines around 4095 and 4640\AA, Ne{\sc ii} 4290\AA, and an
absence of absorption around C{\sc iii}4650\AA.  In addition to
optimizing by interpolation within a grid of precomputed spectra, SFIT
has an option to optimize abundances of individual species with
respect to an observed spectrum for a given model atmosphere. Taking a
model with $T_{\rm eff} = 40\,000$\,K, $\log g = 5.5$ and $n_{\rm He}
= 0.95$, abundances for carbon, nitrogen, silicon and neon were
estimated (Table~\ref{t_params}). If silicon can be taken as a proxy
for metallicity, the star is relatively metal poor, as indicated by
the optimisation described above. No carbon lines are identified,
providing an upper abundance limit. The surface is strongly
CNO-processed (low carbon, high nitrogen), with some evidence of
$\alpha$-capture onto $^{14}{\rm N}$ to form an excess of $^{22}{\rm
  Ne}$. Significant lines at
$\lambda\lambda 3932, 4284$ and  4883\AA\ are
not reproduced in the model fit. $\lambda 3932$\AA\ is likely to be 
due to Ca{\sc ii} and of interstellar origin. $\lambda 4284$\AA\
may be due to S{\sc III} but  the stronger S{\sc III}4254\AA\ is not seen.   
$\lambda 4883$\AA\ could be due to Fe{\sc iii} but, 
again, other Fe{\sc iii} lines are not positively identified.
It is our practice not to accept line identifications without such confirmation.  

\section{{\kep} Photometry}

The detector on board {\kep} is a shutterless photometer using 6 sec
integrations and a 0.5 sec readout. There are two modes of
observation: {\it long cadence} (LC), where 270 integrations are
summed for an effective 28.4 min exposure, and {\it short cadence}
(SC), where 9 integrations are summed for an effective 58.8 sec
exposure.  Gaps in the {\kep} data streams result from, for example,
90$^\circ$ spacecraft rolls every 3 months (called quarters), and
monthly data downloads using the high-gain antenna.

{\kep} data are available in the form of FITS files which are
distributed by the Mikulski Archive for Space Telescope
(MAST)\footnote{http://archive.stsci.edu/kepler}. For LC data each
file contains one observing quarter worth of data whereas for SC data
one file is created per month.  After the raw data are corrected for
bias, shutterless readout smear, and sky background, time series are
extracted using simple aperture photometry (SAP).  The start and end
times of each quarter of {\kep} data used in this study are
shown in Table \ref{log}. SC observations were made in Q3. (We
note that when SC data are obtained, LC data are also produced).

In any wide-angle photometric variability survey, variations in
temperature and, in ground based surveys, seeing and transparency, can
cause correlated variations in the resulting photometric data. The
data from {\kep} is no exception. We show in Table \ref{log} the mean
and standard deviation of the counts of KIC 10449976 for each quarter
of LC data; the latter are shown both before and after removal of the
long-term trend caused by differential velocity aberration.  To remove
systematic trends in the data ({\it e.g.} \citealt{barclay12}) we used
the {\kep} tool {\tt kepcotrend} \citep{still12} which utilises the
correlated global variability of sources in each quarter (this was
done for both SC and LC data). We then filtered the data so that only
points which were flagged {\tt SAP\_QUALITY=0} were kept (this removed
data points which were potentially compromised by events such as solar
flares).  We then normalised the data so that the mean count rate in
each quarter was unity.

\begin{table*}
\begin{center}
\begin{tabular}{lllllrrr}
\hline
Quarter & \multicolumn{2}{c}{Start} & \multicolumn{2}{c}{End} & Mean & Std Dev (I) & Std Dev (II) \\
        & MJD   & UT  & MJD & UT & e$^{-}$ s$^{-1}$& e$^{-}$ s$^{-1}$ & e$^{-}$ s$^{-1}$\\ 
\hline
Q3 (SC) & 55092.722 & 2009 Sep 18 17:05 & 55181.996 & 2009 Dec 17 00:09 & 12963 & 111.9 & 7.1\\
Q5 (LC) & 55275.991 & 2010 Mar 20 23:32 & 55370.660 & 2010 Jun 23 16:05 & 12055 & 190.6 & 8.5\\
Q6 (LC) & 55371.947 & 2010 Jun 24 22:29 & 55461.793 & 2010 Sep 22 19:17 & 11940 & 96.8  & 6.3\\
Q7 (LC) & 55462.672 & 2010 Sep 23 15:53 & 55552.049 & 2010 Dec 22 01:25 & 13379 & 107.8 & 6.8\\
Q8 (LC) & 55567.864 & 2011 Jan 06 20:30 & 55634.846 & 2011 Mar 14 20:32 & 11078 & 12.5  & 6.3\\
Q9 (LC) & 55641.016 & 2011 Mar 21 00:09 & 55738.423 & 2011 Jun 26 10:25 & 11958 & 171.4 & 5.5\\
\hline
\end{tabular}
\end{center}
\caption{The log of {\kep} observations of KIC 10449976. The start and
  end MJD and UT dates are the mid-point of the first and final
  cadence of the LC time series for each quarter respectively. The
  next column shows the mean count rate of KIC 10449976, while the last
  two columns show the standard deviation of points determined from LC
  observations where the effects of long term trends
  are (I) included and (II) removed.}
\label{log}
\end{table*}

\begin{figure*}
\begin{center}
\setlength{\unitlength}{1cm}
\begin{picture}(17,11)
\put(0,0){\includegraphics{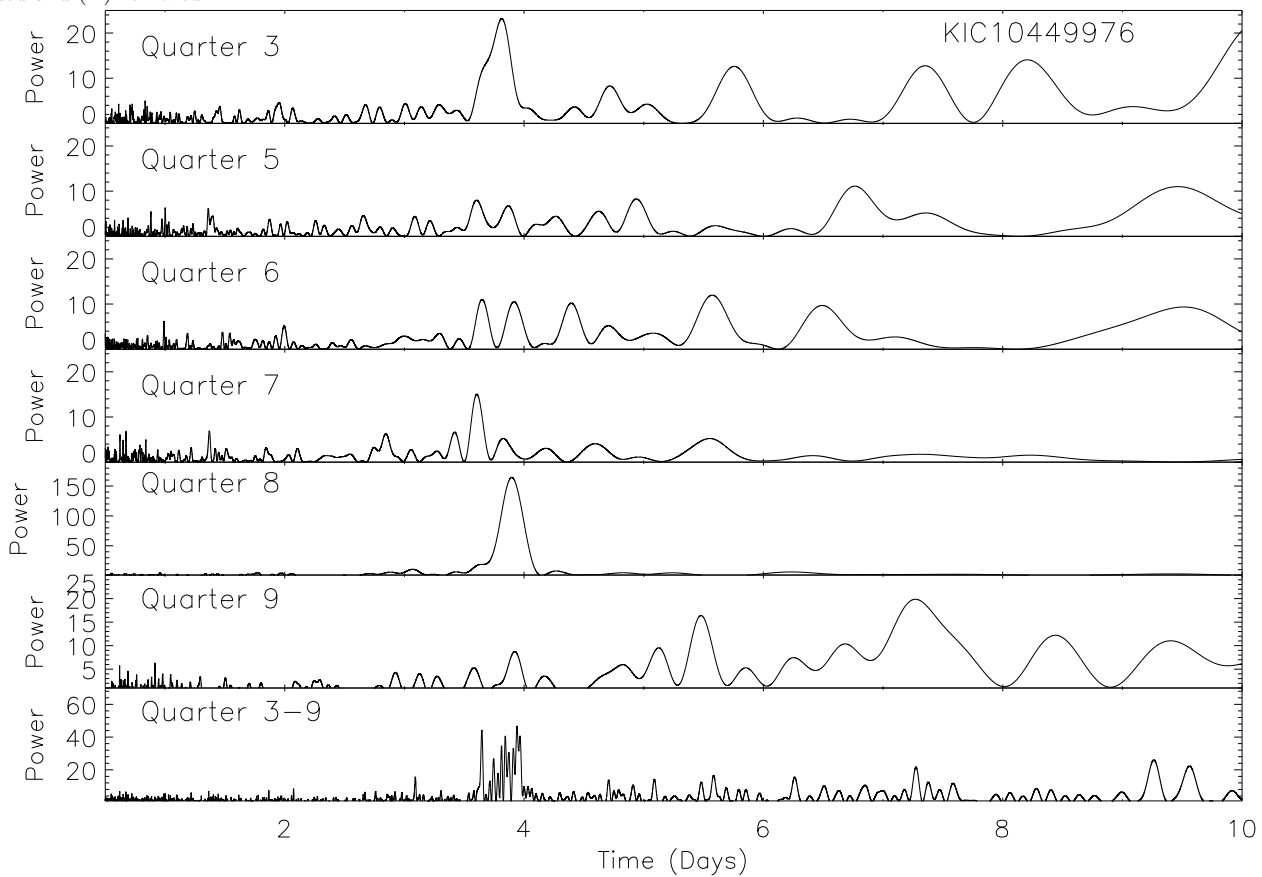}}
\end{picture}
\end{center}
\caption{The Lomb-Scargle power spectra of {\src} photometry taken from different
  quarters after systematic trends have been removed and the data have
  been normalised so that the mean count rate is unity.}
\label{power}
\end{figure*}

\begin{table}
\begin{center}
\begin{tabular}{llc}
\hline
Quarter & Period  & FAP\\
        & (days)  & \\
\hline
Q3 (LC) & 3.81 & --7.5 \\
Q3 (SC) & 3.80 & --11.9 \\
Q5 (LC) & - \\
Q6 (LC) & 5.57 & --2.63 \\
Q7 (LC) & 3.61 & --3.96 \\
Q8 (LC) & 3.90 & --68.97\\
        & 3.65 & --5.75 \\
Q9 (LC) & 7.27 & --6.0 \\
        & 5.48 & --4.5\\
\hline
Q3--9 (LC) & 3.94 & -16.8  \\
           & 3.64 & -15.7  \\   
           & 3.96 & -14.1  \\   
           & 3.84 & -14.1  \\   
           & 3.81 & -11.5  \\   
           & 3.91 & -10.9  \\   
           & 3.87 &  -9.7  \\   
           & 3.74 &  -8.1  \\   
           & 3.78 &  -4.2  \\   
           & 3.09 &  -3.3  \\
\hline
\end{tabular}
\caption{The period (in days) and the logarithm (base 10) of the 
False Alarm Probability (FAP) for periodogram peaks which are 
above the 3$\sigma$ confidence (FAP=--2.5), arranged by quarter.}
\label{power-prob}
\end{center}
\end{table}

We initially searched for periods shorter than 0.5 days using  SC
 data obtained in quarter 3 using the Lomb-Scargle periodogram in
the package {\tt VARTOOLS}  \citep{hartman08b}. There were three
peaks in the power spectra which were detected at a significance
greater than 3$\sigma$ -- 220.7, 196.2 and 252.2 sec. All of these
periods are almost certainly due to spurious signals ({\kep} Data
Characteristics handbook). There is therefore no evidence for a light
modulation of KIC 10449976 with a period less than 12 hrs.

\begin{figure}
\begin{center}
\setlength{\unitlength}{1cm}
\begin{picture}(8,8)
\put(0.5,-1){\includegraphics{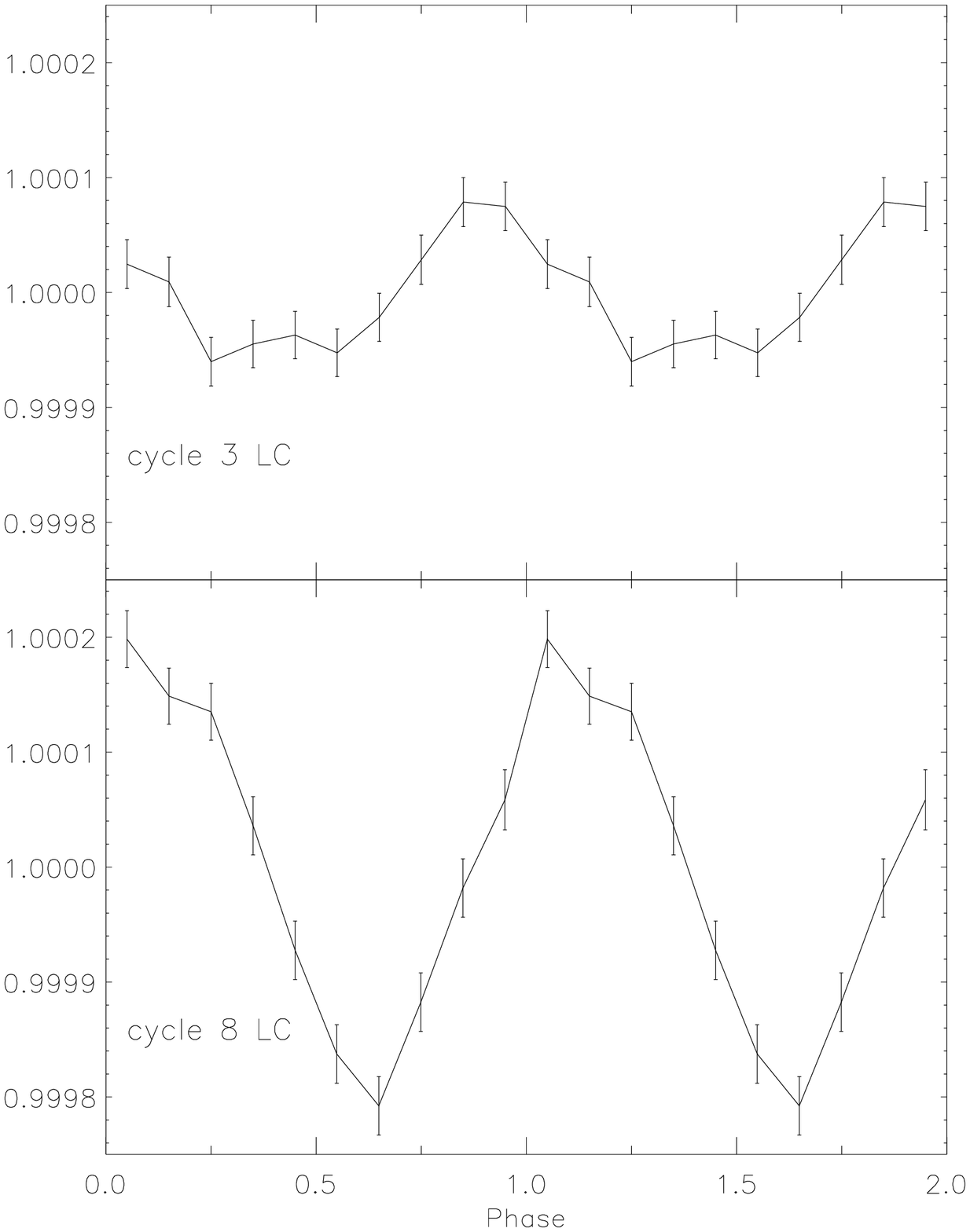}}
\end{picture}
\end{center}
\caption{{\kep} photometry folded on the most promient peak in the power
  spectra (near 3.9 days) in quarters 3 (top) and 8 (bottom).}
\label{fold}
\end{figure}

We then searched for periodic signals in the range 0.5--10 days using
the same procedure as before. We show the resulting power spectra in
Fig.~\ref{power} for each individual quarter and also the power
spectrum from the combined data from quarters 3--9. We list the period
and False Alarm Probability of peaks which are significant at a level
greater than 3$\sigma$ in Table \ref{power-prob}.  We detect
significant peaks in the power spectra at periods between 3--4 days,
the most significant being at 3.9 days.  We folded the data from the
different quarters on the most prominent peaks. We show the folded
light curves from quarters 3 and 8 in Fig.~\ref{fold}. The quarter 3
data folded on the 3.9 day period result in a peak-to-peak modulation
of less than 0.02 per cent; for quarter 8 this becomes 0.04 per cent
(we note that the specific CCD on which {\src} was recorded was
different in quarters 3 and 8). These amplitudes are similar to that
found in the sdB star BD +42$^{\circ}$3250 which has a period of 1.09
day \citep{ostensen10}.

The {\kep} CCDs have a pixel scale of
3.98$^{''}\times3.98^{''}$. We therefore analysed the data on the
pixel level and find that the variability originates from a star which
is on the same pixel as the flux center of KIC 10449976. Using our
images taken using the Wide Field Camera on the INT, we find that the
nearest star ($g$=19.7) to KIC 10449976 is 25.8$^{''}$ distant. We
therefore do not believe that the variability which we detect is due
to blending.

To investigate further the possibility of an instrumental origin, we
obtained data from the {\kep} archive of 16 stars which are within 10
arcmin and 0.5 mag and observed in the same quarters
as {\src}. Four of these stars (KIC 1044981, 10450110, 10515199 and
10580086) showed a clear modulation on a dominant period of 6--10 days
in their {\kep} light curve. None showed a modulation on a period
close to 3.9 days. Several more stars showed evidence for less
significant periods close to 8 days. The remaining stars showed no
evidence for a period seen in every quarter or at a period of 3.9 days. 
Although we cannot rule out the possibility that the periods noted in Table
\ref{power-prob} are due to an instrumental effect, there are good
grounds to conclude that they are astrophysical in origin.

\begin{figure*}
\begin{center}
\setlength{\unitlength}{1cm}
\begin{picture}(16,10)
\put(0,-9){\includegraphics{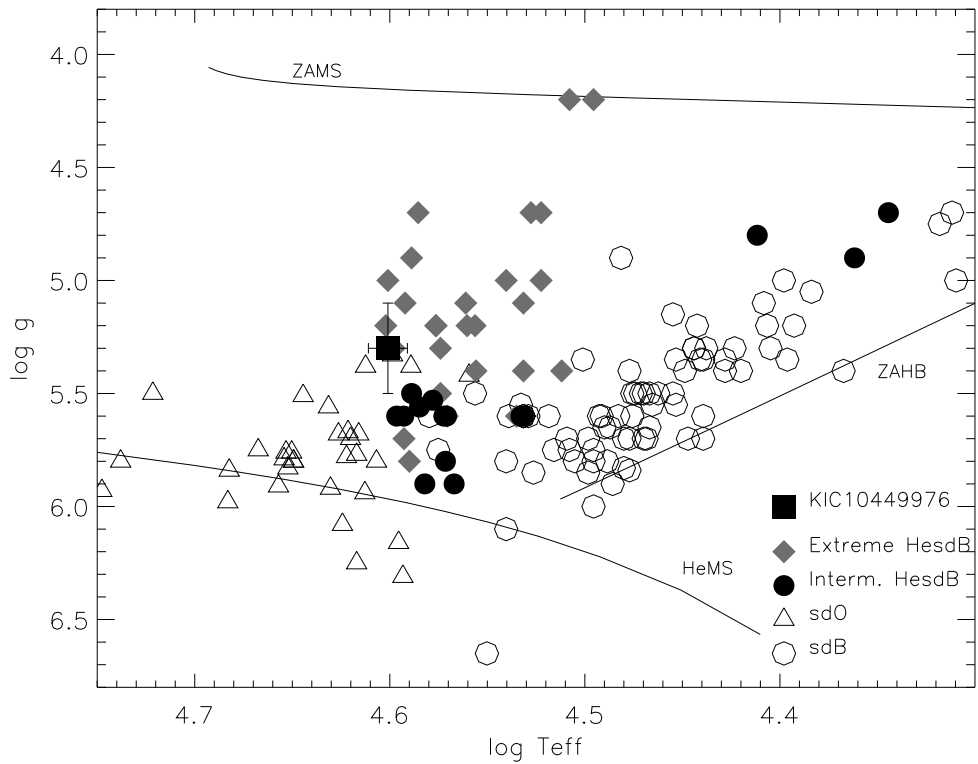}}
\end{picture}
\end{center}
\caption[$g- T_{\rm eff}$ digram for hot subdwarfs]{The
  surface-gravity -- effective-temperature diagram for hot subdwarfs
  showing the positions of normal sdB and sdO stars (open symbols),
  intermediate helium-rich subdwarfs (filled circles), extreme
  helium-rich subdwarfs (filled diamonds) and KIC 10449976 (filled square). 
  Approximate
  positions of the zero-age hydrogen main sequence (ZAMS), the helium main
  sequence (HeMS)  and the zero-age extended horizontal branch (ZAHB) 
  are also indicated. Data for
  individual stars are taken from \citet{naslim10}, \citet{naslim11},
    \citet{stroeer07}, \citet{ahmad03}, and \citet{edelmann03}.}
\label{f_gteff}
\end{figure*}

\section{Discussion}

With $n_{\rm He} > 0.90$ and its position on the $g - T_{\rm eff}$
diagram (Fig.~\ref{f_gteff}), KIC 10449976 is well-established as an
extreme helium-rich subdwarf (EHe-sd), and amongst the hottest of
its class.  \citet{zhang12a} argue that
the hottest EHe-sds are also the most massive, having $M>0.70 {\rm
  M_{\odot}}$, although they also point out that, unlike KIC 10449976,
these are also predominantly carbon-rich.

A number of questions are posed by the {\it Kepler} photometry.
Foremost is whether the variation is genuine and intrinsic to the
star. We have examined whether the variation could be due to an
instrumental effect and, while we cannot rule out this possibility, 
we believe it more likely to be astrophysical in origin and
proceed on this basis. The second question concerns the cause of the
variability. The folded light curve points to a sinusoidal variation
which could originate either in pulsation, in reflection from a
companion, or from the rotation of a non-spherical star (ellipsoidal
variation).  All three present difficulties.

Many hot subdwarfs do pulsate
\citep{kilkenny97a,green03,jeffery05.india}.  For a star with the
dimensions of KIC 10449976, the fundamental radial mode would have a
period of the order of 200\,s. Even high-order gravity modes are
unlikely to have periods in excess of a few hours.  The pulsating
helium-rich subdwarf LS IV-14 116 shows up to six g-mode oscillations
with periods between 1953 and 5084\,s
\citep{ahmad05,green11,jeffery11.ibvs}. Pulsation periods more than 20
times this value would only be expected in stars where the radius is
some 10 times larger, or the gravity was 1 dex lower, a possibility
that is ruled out by the spectroscopy.

Hot subdwarfs in HW\,Vir-type close binary systems show light
variations due to reflection from an M-dwarf companion ({\it e.g.}
\citealt{lee09}).  Reflection-effect light variations should be
strictly periodic over the time interval covered by the {\it Kepler}
observations.  Supposing that observational errors account for the
quarterly variation in apparent period, a simple test calculation
involving only assumed masses for the helium star and its companion
({\it e.g.} $M_1=0.7 {\rm M_{\odot}}$ and $M_2=0.38 {\rm M_{\odot}}$),
the albedo of the companion ({\it e.g.} $A=0.9$), the inclination
({\it e.g.} $i=45^{\circ}$) and the measured period (3.9\,d) can be
used to estimate the amplitude of the orbital variation in both total
light (0.016\%) and projected helium-star radial velocity ($50\,{\rm
  km\,s^{-1}}$).  In order to simulate a system in which the projected velocity
amplitude is less than 50 km s$^{-1}$ and the light amplitude is
greater than 0.02 per cent, we require a high albedo ($A>0.8$), a low
inclination ($i<50^{\circ}$) and, with 
$M_1 \geq 0.7\,{\rm M_{\odot}}$ for example, $0.20 \leq M_2/{\rm M_{\odot}} \leq 0.33 $. 
These are not unreasonable values; the principal remaining argument against a
reflection-effect origin is the lack of stability in the apparent
period.

Following a similar argument and making reasonable assumptions about
masses and radii, the maximum contribution of any ellipsoidal
contribution to the total light can be computed \citep{morris85} and
is found to be negligible ($<10^{-4}\%$) in this case.

A further and quite realistic possibility is the existence of spots on the stellar
photosphere, as found in certain helium-rich B stars on the main
sequence; {\it e.g.}  $\sigma$\,Ori\,E \citep{greenstein58}. Such
spots are usually associated with the poles of strong magnetic fields
and surface chemical inhomogeneity; surface brightness variations are
tied to the rotation of the star's magnetic axis \citep{townsend05}.
Moreover, spots are relatively short lived and have been observed
to give a {\kep} signature similar to that reported here (Balona 2013, in preparation). 
In this case, the constraint of constant period and amplitude would be
less strong, but is still not negligible, since the spots are locked
in position by the magnetic field. 

Finally, intrinsic variability in an unresolved and fainter close companion
cannot be ruled out, although any mechanism responsible would 
be subject to a similar commentary to that outlined above. There is
no evidence for such a companion in the spectrum, but either an M dwarf
or a white dwarf would not be detected spectroscopically.

It is emphasized that all of the above interpretations for the
apparent light modulation in {\src} are, at this stage, conjecture.
The discovery of an extremely helium-rich subdwarf in a close binary
would have profound consequences for a proposed white-dwarf merger
origin \citep{zhang12a}.  Further observations are necessary to
investigate the long-term stability of the period, and to increase the
precision of the radial-velocity measurements.

\section{Conclusion}

\src\ is a blue star in the {\kep} input catalogue which shows 
small-amplitude light variations with a period of approximately 3.9\,d. Our
spectroscopic classification identified it as a helium-rich subdwarf, a
conclusion which is confirmed by higher-resolution spectroscopy. The
latter shows the helium abundance to be greater than 90\% by numbers,
and that the star is silicon-poor, nitrogen-rich and neon-rich. Carbon
was not detected. On the basis of effective temperature and surface
gravity, \src\ should have $M>0.7 {\rm M_{\odot}}$, but the high
nitrogen/carbon ratio implies $M<0.7 {\rm M_{\odot}}$, suggesting a
mass close to this boundary.  An upper limit of $50\pm20$\,km\,s$^{-1}$ in
radial-velocity amplitude places constraints on any putative binary
companion.  The absence of long-term stability in the light variation
is also a challenge for a binary interpretation, although the
amplitudes could be consistent with a reflection-effect solution. The
period is far too long to be due to pulsation. Starspots remain a plausible
interpretation.  Further high-quality
radial-velocity and photometric studies will be required to establish
the authenticity and cause of the light variations.

\section*{Acknowledgments}

The Armagh Observatory is supported by a grant from the Northern
Ireland Dept. of Culture Arts and Leisure.  The INT and WHT are
operated on the island of La Palma by the Isaac Newton Group in the
Spanish Observatorio del Roque de los Muchachos of the Instituto de
Astrofísica de Canarias. We thank the staff for their support.  This
paper includes data collected by the {\kep} mission. Funding for the
Kepler mission is provided by the NASA Science Mission
Directorate. Some of the data presented in this paper were obtained
from the Mikulski Archive for Space Telescopes (MAST). STScI is
operated by the Association of Universities for Research in Astronomy,
Inc., under NASA contract NAS5-26555. Support for MAST for non-HST
data is provided by the NASA Office of Space Science via grant
NNX09AF08G and by other grants and contracts.

\bibliographystyle{mn2e}
\bibliography{mnemonic,ehe}

\label{lastpage}

\end{document}